\documentclass[aps,prl,groupedaddress,nofootinbib,notitlepage,showpacs,floatfix,twocolumn]{revtex4-1}

\usepackage{graphicx,graphics,epsfig,subfigure,times,bm,bbm,amssymb,amsmath,amsfonts,mathrsfs}
\usepackage[matrix,frame,arrow]{xypic}
\usepackage[pdfstartview=FitH]{hyperref}
\usepackage{hypernat}
\usepackage{graphicx}
\usepackage{epstopdf}
\usepackage[normalem]{ulem}

\newcommand{\beq}{\begin{equation}}
\newcommand{\eneq}{\end{equation}}
\newcommand{\beqnn}{\begin{equation*}}
\newcommand{\eneqnn}{\end{equation*}}
\newcommand{\beqy}{\begin{eqnarray}}
\newcommand{\eneqy}{\end{eqnarray}}
\newcommand{\beqynn}{\begin{eqnarray*}}
\newcommand{\eneqynn}{\end{eqnarray*}}

\newcommand{\ket}[1]{ | #1\rangle}
\newcommand{\bra}[1]{\langle #1 | }

\newcommand{\ignore}[1]{}
\newcommand{\vp}{\vec{p}}
\newcommand{\Or}{\mathcal{O}}

\begin{document}

\author{Silvano Garnerone$^{(1,2,5)}$, Paolo Zanardi$^{(2,5)}$, Daniel A. Lidar$^{(2,3,4,5)}$}
\affiliation
{$^{(1)}$Institute for Quantum Computing,
University of Waterloo, Waterloo, ON N2L 3G1, Canada\\ 
Departments of $^{(2)}$Physics \& Astronomy, $^{(3)}$Electrical Engineering, $^{(4)}$Chemistry, and $^{(5)}$Center for Quantum Information Science \&
Technology,
 University of Southern California, Los Angeles, CA 90089, USA}

\title{Adiabatic quantum algorithm for search engine ranking}

\begin{abstract}
We propose an adiabatic quantum algorithm 
for generating a quantum pure state encoding
of the PageRank vector, 
the most widely used tool in ranking the relative importance of internet pages. 
We present extensive numerical simulations which provide evidence that 
this algorithm {can} prepare the quantum PageRank state in a time which, {on average}, scales polylogarithmically in the number 
of webpages. {We argue that the main topological feature of the underlying web graph allowing for such a scaling is the out-degree distribution.}
The top ranked $\log(n)$ entries of the quantum PageRank state
can {then} be estimated with a polynomial quantum speedup.
Moreover, the quantum Page{R}ank state can be used in ``q-sampling'' protocols 
for testing properties of distributions, which require exponentially fewer measurements 
than all classical schemes designed for the same task. 
This can be used to decide 
whether to run a classical update of the PageRank.
\end{abstract}

\maketitle

{\it Introduction}.---%
Quantum mechanics provides computational resources that 
can be used to outperfom classical algorithms \cite{NiCh}. 
Problems for which a polynomial or exponential quantum  speed-up 
is achievable have been sought in quantum computation since its inception, 
and their ranks are swelling slowly \cite{Bacon:2010}. 
{Yet, while ranking the results obtained in response to a user query is 
one of the most difficult tasks in searching the web \cite{Manning:2008:IIR:1394399}, 
so far no efficient quantum algorithms have been proposed for this task \cite{old-PR-comment}.}

Here we present an adiabatic quantum algorithm {\cite{FaGoGu}}
which prepares 
a state 
containing the same ranking information as the 
PageRank vector. The latter is 
{a central} tool   
in data mining and information retrieval, 
at the heart of the success of the Google search engine 
\cite{Pagerank,Manning:2008:IIR:1394399,citeulike:796239,Bon,doi:10.1080/15427951.2005.10129098}.
{The best available classical} algebraic and {Markov Chain} Monte Carlo {(MCMC)} techniques used to evaluate the full PageRank vector 
require a time which scales 
as $O(n)$ and $O(n\log n)$, respectively,
where $n$ is the number of pages, i.e., the size of the web-graph. 
{We investigate the size of the gap {of the adiabatic Hamiltonian} numerically}
using a wide range of web-graph sizes ($n \in \{2^2,\dots, 2^{14}\}$), {and present evidence}
that our quantum algorithm prepares the PageRank state in a time which scales {on average} as $O[\textrm{polylog}(n)]$.
We argue that while extraction of the full PageRank vector cannot in general be done more 
efficiently than when using the aforementioned classical algorithms, there are 
{particular graph-topologies and}
specific tasks of relevance in the 
use of search engines for which the quantum
algorithm, combined 
with other known quantum protocols \cite{Aharonov:2003:AQS:780542.780546, 5773032, Brassard02quantumamplitude, PhysRevLett.87.167902},  {may} provide a polynomial, or even
exponential speedup. {We discuss the underlying graph structure which we believe is responsible for this potential speedup, and provide evidence that it is the power law distribution of the out-degree nodes that plays the key role. A proof of this fact would be very interesting.}

{\it Model of the web-graph.---}%
{The PageRank algorithm, introduced by Brin \& Page \cite{Pagerank}, is probably the most prominent ranking measure 
using 
{the query-independent hyperlink structure of the web.}
The PageRank vector is the principal eigenvector of the 
so-called Google matrix, which encodes the structure of the web-graph via its adjacency matrix. The humongous size of the World Wide Web (WWW), with its ever growing number of pages and links,
makes the evaluation of the PageRank vector one of the most demanding 
computational tasks ever \cite{doi:10.1080/15427951.2005.10129098}.}
In practice 
{PageRank} is evaluated over real data providing the 
structure of the actual WWW. 
On the other hand the use of models of the web-graph 
has proved to be useful in testing new ideas concerning 
structure measures and dynamical properties of the web \cite{Bon}. 
To accurately capture the WWW graph a good candidate model network should be 
(i) sparse (the number of edges is proportional to the number of nodes), (ii) small-world (the network diameter
scales logarithmically in the size of the network), and (iii) scale-free (the in- and out-degree probability distributions obey a power law).
To analyze the scaling properties 
of our algorithm we used two well known models of the web-graph: the preferential 
attachment model 
\cite{BaAl}, and the copying model \cite{KlKuRa}. 
These models are based on two different network evolution mechanisms, both of which 
yield sparse random graphs with small-world and scale-free (power-law) features.

We implemented a version \cite{BoRiSp} of the 
preferential 
attachment model that provides a scale-free network with
$N(d)\propto d^{-3}$, where $N(d)$ is the number of nodes of degree $d$. 

The copying model \cite{KlKuRa} improves upon the preferential attachment model by exploiting only local structure to 
generate a power-law degree distribution, and providing for random graphs with $N(d)\propto d^{(2-p)/(1-p)}$, where $p$ is a probability \cite{sup-mat}.

{\it Google matrix and PageRank.---}%
PageRank can be seen as the stationary distribution 
of a random walker on the web-graph{, which spends its time on each page in proportion to the relative importance of that page} \cite{citeulike:796239}. 

To model this
define the 
transition matrix $ P_1 $ associated with the adjacency matrix $ A $ of the graph
\begin{equation}
P_1(i,j)=
\left\{ \begin{array}{ll}
 1/d(i) & \mbox{if $(i,j)$ is an edge of $A$};\\
 0 & \mbox{else},\end{array} \right.
\end{equation}
where $ d(i) $ is the out-degree of the $i$th node. 

Since the out-degree of a node might be $0$, a 
walker that follows only
links can become trapped in a
node with no out-links. 
Equivalently, if $P_1$ has a row of all $0$'s then it is not stochastic.
To overcome this problem 
one modifies $ P_1 $  by replacing 
every zero row with the vector $ \vec{e}/n $ whose entries are all $ 1/n $. 
Call this new stochastic matrix $ P_2 $. 
However, there is still the possibility of ``importance sinks,'' meaning subgraphs with in-links but no out-links, i.e., $P_2$ needs to be made irreducible \cite{com-irred}. 
To accomplish this one defines the Google matrix $ G $ as
\begin{equation}
G:=\alpha 
P_2^T + (1-\alpha)E,
\end{equation}
where  $ E \equiv \ket{\vec{v}} \bra{\vec{e}}$.

The ``personalization vector" $ \vec{v} $ is {a probability distribution with all positive entries}; 
the typical choice is $\vec{v}=\vec{e}/n $.
The parameter $\alpha$ is the probability that the 
{walker} follows the link structure of the  web-graph at each step, rather than hop randomly 
between graph nodes according to $ \vec{v} $. 
Google reportedly uses $ \alpha=0.85 $, which we also 
use in this work.
The matrix $ E $ makes $ G $ irreducible and aperiodic,
and hence the Perron-Frobenius theorem 
ensures the existence of a unique eigenvector with all positive entries
associated to the maximal eigenvalue $1$. This 
eigenvector is precisely the PageRank $ \vp $ \cite{citeulike:796239}. Moreover, 
the modulus of the second eigenvalue of $G$ is upper-bounded by $\alpha$ \cite{gap-bound}. 
This is important for the convergence of the power method, the standard computational 
technique employed to evaluate $ \vp $. It uses the fact that for any 
probability vector $ \vp_0 $ 
\begin{equation}
\vec{p}=\lim_{k\rightarrow \infty} G^k \vec{p}_0.
\end{equation}
The power method computes $\vp$ with accuracy  $\nu$ in a time $O[sn\log({\nu})/\log(\alpha)]$,
where $s$ is the sparsity of the graph (maximum number of non-zero entries per row of the adjacency matrix). 
The rate of convergence is determined by $\alpha$. 
The other technique 
used in the evaluation of  
PageRank is MCMC,
where a direct simulation of rapidly mixing random walks is used 
to estimate the PageRank at each node. 
The typical running time is 
$O[n\log(n)]$
\cite{Bahmani:2010:FIP:1929861.1929864}. 

{\it Adiabatic quantum computation.---}%
Even though {classical} PageRank computation time scales {modestly}
with the problem size $n$, in practice its evaluation for the actual WWW already takes weeks, 
a time which can only be expected to grow if current computational methods remain the norm, 
given the rapid pace of expansion of the web. Furthermore, it is often desirable to have multiple 
personalization vectors, which means that more than one PageRank needs to be evaluated for each WWW graph instance. 
Considering also the fact that 
the web-graph is an evolving dynamic entity, it is clear that it is 
important to speed up the computation of the PageRank in order to 
provide up-to-date results from the ranking algorithm. 

We now show how adiabatic quantum computation (AQC) \cite{FaGoGu,Farhi20042001,PhysRevA.65.042308,PhysRevLett.101.170503,PhysRevLett.106.050502}
might be able to help in the optimization of  the resources 
needed to provide an up-to-date PageRank. 

Small-scale experiments  {with the potential to pave} the way toward laboratory realization of AQC, involving $8$ superconducting flux qubits, have recently been reported \cite{DWave-Nature}. In AQC one 
encodes the solution to a difficult problem in the 
ground state of a related problem Hamiltonian $H^{(p)}$. 
The latter is arrived at by slowly modifying an 
initial Hamiltonian $H^{(i)}$, for which the ground state is---by construction---easy to obtain. 
The adiabatic evolution is generated by  
$H(s)=(1-s)H^{(i)}+sH^{(p)}$.
If the modification from the initial to the final Hamiltonian is done slowly 
enough, and the parameter $s(t):0\mapsto 1$ has a smooth time dependence, where the time $t\in [0,T]$,
then the quantum adiabatic theorem 
guarantees that the state of the system will be the ground state 
for all $t$ with high probability \cite{Teufel:book}. 
More precisely, in order for the 
final system state $|\psi(T)\rangle = \mathcal{T} e^{-i \int_0^T H{[}s(t){]}dt}|\psi(0)\rangle $ to have fidelity 
\begin{equation}
f:=|\langle \psi(T)|\pi\rangle | \gtrsim 1-\eta^a
\label{eq:f}
\end{equation}
with respect to the the desired ground state $ |\pi\rangle $ of $H^{(p)}$, the total adiabatic evolution time should satisfy
\begin{equation}
T \gtrsim {a}\frac{\Lambda^{b-1}}{\eta \delta^{b}},
\label{Eq:bound}
\end{equation}
where  $ \Lambda=\max_s \Vert dH/ds \Vert$ 
(the norm is the largest eigenvalue) and $ \delta=\min_s \Delta(s)$, where $ \Delta(s) $ is the instantaneous energy gap of $ H(s) $ between the ground and first excited state.
The values of the integer exponents $a$ and $b$ in Eqs.~\eqref{eq:f} and \eqref{Eq:bound} {depend upon the differentiability and analyticity properties of $H(s)$, and the boundary conditions satisfied by its derivatives;
typically $b \in \{1,2,3\}$ \cite{Jansen:06}, while $a$ can be tuned between $1$ and arbitrarily large integer values, equal to the number of vanishing derivates of $H(s)$ at the boundaries $s=0$ and $s=1$} \cite{ad}.

{\it Adiabatic quantum PageRank algorithm.---}
Since $G$ is not reversible we cannot directly apply the standard technique of mapping it to a discriminant matrix 
without {\textit{a priori} knowledge of the stationary state}
\cite{Szegedy:04,Aharonov:2003:AQS:780542.780546,Krovi:10}. Instead, let us consider the following 
non-local problem Hamiltonian associated with
a generic Google matrix $ G $ 
(note that we use $H$ and $h$ for
local and non-local Hamiltonians, respectively):
\begin{equation}
h^{(p)}=h(G)\equiv\left(\mathbb{I}-G \right)^{\dagger} \left( \mathbb{I}-G \right).
\label{Eq:hp}
\end{equation}
Since $ h(G) $ is positive semi-definite, and $1$ is the maximal
eigenvalue of $ G $ associated with $\vp$, it follows 
that the ground state of $ h(G) $ is given by $ |\pi\rangle \equiv \vp/\Vert \vp \Vert_2 $. 
The initial Hamiltonian has a similar 
form, but it is associated with the Google matrix $ G_c $ of the 
complete graph \cite{pagerank-comment1}
\begin{equation}
h^{(i)}=h(G_c)\equiv\left(\mathbb{I}-G_c \right)^{\dagger} \left( \mathbb{I}-G_c \right).
\label{Eq:hi}
\end{equation}
The ground state of $ h^{(i)} $ is $ |\psi(0)\rangle = \sum_{j=1}^n|j\rangle/\sqrt{n} $, 
{a fully delocalized, uniform quantum superposition state. The basis vectors $|j\rangle $ span the $n$-dimensional Hilbert space of $\log n$ qubits.}
The interpolating adiabatic Hamiltonian is
\begin{equation}
h(s)=(1-s)h^{(i)} + s h^{(p)}.
\label{Eq:hs}
\end{equation}
Equations~\eqref{Eq:hp}-\eqref{Eq:hs} completely 
characterize the adiabatic quantum PageRank algorithm, apart from the interpolation function $s(t)$, which can be optimized using differential geometric or variational methods to simultaneously minimize the adiabatic evolution time $T$ and the adiabatic error $ \varepsilon:=\sqrt{1-f^2} $ \cite{PhysRevLett.103.080502,PhysRevA.82.012321,PhysRevA.82.052305}. 
By simulating the dynamics generated by $ h(s) $
we can estimate the 
parameters in Eq.~\eqref{Eq:bound} {\cite{sim-comment}}. 

\begin{figure}
\includegraphics[scale=0.45]{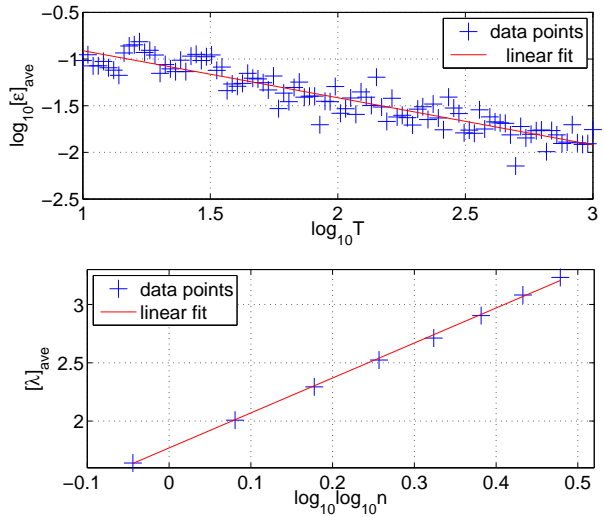} 
\caption{(color online) 
Top panel: The typical adiabatic error $ [\varepsilon]_\textrm{ave}$ 
scales approximately as $T^{-0.48}$. Results are for a system of size $n=16$ (we checked different sizes obtaining similar results), averaged over 100 random web-graph realizations. Bottom panel:  
 {$[\lambda]_\textrm{ave}$ scales as $\log \log n$, with a prefactor which is 
approximately 3}. Results were averaged over $1000$ random web-graph realizations. See text for more details.}
\label{Fig:fig1}
\end{figure}

\begin{figure}
\includegraphics[scale=0.4]{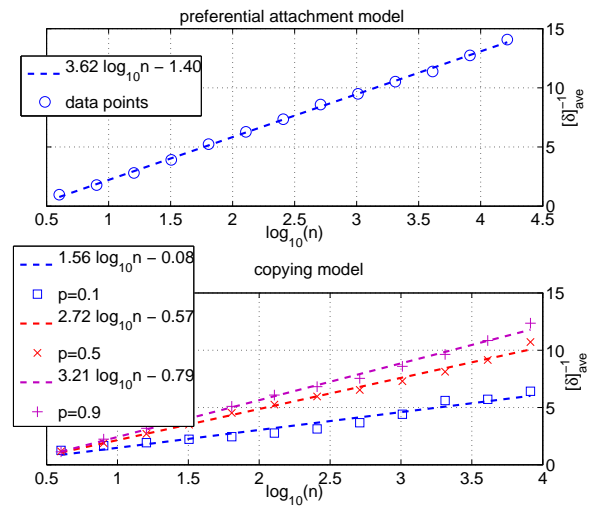} 
\caption{
(color online) Scaling of the {inverse of the} average 
minimum gap{, $1/[\delta]_{\rm ave} $,} for the preferential attachment model (top panel), 
and copying model (bottom panel). 
{We checked numerically that the same logarithmic scaling holds for the 
averaged inverse gap $ [1/\delta]_{\rm ave} $, with the latter slightly larger than $ 1/[\delta]_{\rm ave} $ for all graph sizes $n$ \cite{sup-mat}. 
Note that for the copying model the parameter $p$, relative to the in-degree distribution, affects only the prefactor of the scaling.}
}
\label{Fig:fig2}
\end{figure}

{\it Simulation results}.---%
Figures \ref{Fig:fig1} and \ref{Fig:fig2} summarize our numerical simulations 
on the USC high-performance cluster \cite{cluster}. Figure \ref{Fig:fig1} shows the results for the 
preferential attachment model,
providing information on the adiabatic error $ \varepsilon$ and the scaling 
of $ \lambda\equiv \|dh/ds\| = \Vert {h}^{(p)}-{h}^{(i)}\Vert $ [corresponding to the numerator in Eq.~\eqref{Eq:bound}],
with respect to the number of web-graph nodes. In these simulations we made no attempt to minimize the error by optimizing $s(t)$. 
From the upper panel we can conclude that the adiabatic run-time $T$ scales as the inverse square of the adiabatic error $ \varepsilon$. The bottom 
panel shows the ensemble average of $ \lambda$.
The fit clearly shows that {for the preferential attachment model} {$\lambda$} 
exhibits a double logarithmic scaling as a function of $ n $.
We checked numerically that similar results hold 
also for the copying model (not shown). 

Figure~\ref{Fig:fig2} displays the scaling of the minimum gap with respect to 
system size, also averaged over $1000$ 
random web-graph realizations. The top panel displays the results for the preferential attachment 
model. The bottom panel is for the copying model, for which we 
considered different values of the parameter $ p $. 
{In both models} {the random graphs} {were} {generated so that they have both in- and out-degree power-law distributions.}  {More specifically, we mixed (i.e., added the adjacency matrices of) 
graphs $\mathcal{G}_A$, with only in-degree power-law distributions, with graphs 
$\mathcal{G}_B$ with only out-degree power-law distributions. 
For the simulations reported here, 
the maximum out-degree for $\mathcal{G}_B$ is approximately $3$ times greater than the maximum in-degree 
for $\mathcal{G}_A$}. Our
simulation results, which cover nearly four orders of magnitude of graph sizes, indicate
that, {for the class of graphs we have considered,} the inverse of the 
average gap is proportional to $\log(n)$.

Putting together the above observations, 
namely that for a typical graph instance  
$\lambda \sim {\rm poly}(\log\log n)$, $\delta \sim 1/{\rm poly}(\log n)$, 
$T \sim \varepsilon^{-c}$ (with $ c \approx 2 $, see 
Figure \ref{Fig:fig1}), we can 
conclude {from Eq.~\eqref{Eq:bound}} that the typical run-time of 
the adiabatic quantum PageRank algorithm scales  as
\begin{equation}
T\sim \varepsilon^{-2}(\log\log n)^{b-1} (\log n)^{b},
\label{Eq:compcomp}
\end{equation}
where $ b $ is some small {positive} integer that depends on the details of the network topology 
(see Fig. \ref{Fig:fig2}). We checked this result by simulating the adiabatic
evolution of the system allowing for a run-time {$ T=\epsilon^{-2}(\log\log n)^{b-1}(\log n)^b $}, 
with both $b=2$ and $b=3$ for small graphs (up to 20 nodes), with
a fixed small {$ \epsilon $}.
For each evolving random graph we found that the final calculated 
adiabatic error {$\varepsilon$} is always upper bounded by {$ \epsilon $}.

{\it Mapping to a local Hamiltonian}.---%
Since the 
Google matrix $ G $ is not sparse, the physical implementation 
of the {$\log n$-qubits} Hamiltonian in Eq.~\eqref{Eq:hs} can, in general, require many-body interactions with arbitrarily high locality. This problem is similar to one that arises, e.g., in the quantum adiabatic implementation of Grover's search algorithm \cite{PhysRevA.65.042308}. A general technique to overcome the non-locality problem is the use of so-called perturbation gadgets, which requires the introduction of ancillary qubits \cite{Jordan:08}. However, a more direct alternative is to 
map the dynamics generated by Eq.~\eqref{Eq:hs} from the 
$ n $-dimensional Hilbert space into the $n$-dimensional single particle excitation subspace of an 
effective $ 2^n $-dimensional Hilbert space with $ n $ qubits. 
This correspondence has been used 
recently in a different context 
to study the quantum dynamics of biomolecular systems \cite{lhc}, 
and it has also been considered from an experimental 
perspective \cite{alan}. 
The new effective adiabatic Hamiltonian is given by
\begin{equation}
H(s)=\sum_{i=1}^n h(s)_{ii} \sigma_i^+ \sigma_i^- 
+ \sum_{i<j}^{n} h(s)_{ij} \left(\sigma_i^+ \sigma_j^- 
+ \sigma_j^+ \sigma_i^-\right),
\label{Eq:bigH}
\end{equation}
where $h(s)_{ij}$ is the $(i,j)$th matrix element of $h(s)$ as given in Eq.~\eqref{Eq:hs}, and $\sigma_i^{\pm}$ is the Pauli raising or lowering matrix for the $i$th qubit (or web-graph node) \cite{stoq}. 
The spectral properties of $ H(s) $ {in the single particle excitation subspace} are the same 
as those of $h(s)$ {\cite{sup-mat}}. This implies that the 
estimate \eqref{Eq:compcomp} also holds 
for $ H(s) $, and hence one could envision programming $H(s)$ of Eq.~\eqref{Eq:bigH} onto physical systems such as excitonic quantum dots or flux qubits, where two-qubit
coupling has been shown to be sign- and magnitude-tunable \cite{Hime01122006,PhysRevLett.98.057004,PhysRevB.82.024511}. {Provided this programming step can be executed in time
at most $O(\log n)$,
updating the matrix elements $h(s)_{ij}$
is efficient \cite{DWave-couplers-update}.}

At the conclusion of the adiabatic evolution generated by the Hamiltonian in Eq.~\eqref{Eq:bigH}, the 
PageRank {vector $\vec{p}=\{p_i\}$} is encoded into the {quantum PageRank} state $\ket{\pi}=\sum_{i=1}^n \sqrt{\pi_i}
\ket{i}$ of an $n$-qubit system, where 
{$ |i\rangle $ is the vector with $1$ in the $i$th entry, and $0$'s in all the others.}
The probability of finding the only allowed excitation 
at site $i$ is $\pi_i = p_i^2/\Vert \vec{p} \Vert _2^2$. One can estimate ${\pi_i}$ by repeatedly 
sampling the expectation value of the operator $ \sigma^z_i $ in the final state.  
The number of measurements $M$ needed to estimate $\pi_i$ is {given} by the Chernoff-Hoeffding 
bound \cite{citeulike:3392582}, allowing us to approximate {$\pi_i$} with an additive error $e_i$ 
and with $M={\rm poly}(e_i^{-1})$.
We now discuss tasks for which the quantum ranking algorithm offers a speedup.

\textit{Ranking the top}.---The fact that the 
amplitudes of the quantum PageRank state are $\{\sqrt{\pi_i}\}$, 
rather than $\{\sqrt{p_i}\}$, 
is in fact a virtue: we can show that  $\forall i$
the total quantum cost is $O[n^{2\gamma_i-1}\textrm{polylog}(n)]$ 
for estimating the rank $\pi_i$
with additive error $e_i\sim\pi_i$, while the corresponding 
classical cost is at best $O[n^{\gamma_i} \log(n)]$ \cite{gain}. 
Thus for this task there 
is a polynomial quantum speedup whenever $\gamma_i<1$; 
our simulations show that this is indeed the case for the top-ranked $\log(n)$ pages. 

{\textit{Comparing successive PageRanks}.---}%
Another context for useful applications is ``q-sampling" \cite{Aharonov:2003:AQS:780542.780546}.
Since the classical PageRank algorithm is so costly when applied to the WWW,  
one would like to develop criteria for when to run it, e.g., after a
relevant perturbation to the graph. 
The adiabatic quantum algorithm 
can provide, in time $O[{\textrm{polylog}}(n)]$, the pre- and post-perturbation
states $\ket{\pi}$ and $\ket{\tilde{\pi}}$ as input to a quantum circuit 
implementing the SWAP-test \cite{PhysRevLett.87.167902}.
To obtain an estimate of the fidelity $|\bra{\pi}\tilde{\pi}\rangle|^2$ we need to measure an ancilla $O(1)$ times, the number depending only on the desired precision.
Whenever some relevant perturbation 
of the previous 
quantum PageRank state is observed, one can decide to run 
the classical algorithm again to update the 
classical {PageR\nolinebreak[4]ank}. 
Deciding whether two probability distributions---one of which is known---are close, classically requires approximately $\sqrt{n}$
samples \cite{5773032,10.1109/SFCS.2001.959920}. 
Related quantum algorithms for testing properties of distributions \cite{Valiant:2008:TSP:1374376.1374432} 
have recently been proposed and analyzed \cite{5773032}.

{\it Discussion}.---%
Why do we observe a ``large" gap that scales as $O(1/{\rm poly}(\log n))$? 
The out-degree distribution seems to be the key feature activating the polylogarithmic behavior \cite{sup-mat}. In support of this claim we have also analyzed two other classes of random graphs: 
one with only in-degree power-law distribution, the other with only out-degree power-law distribution. 
In the former we found that the average inverse gap scales polynomially in the system size (``small" gap), while in latter we found the ``large" gap, polylogarithmic scaling. On the other hand when the out-degrees are equal to the in-degrees (as for undirected graphs) the gap scaling is again polynomial. The scaling for intermediate cases is determined by the presence or absence of sufficiently many nodes linking to a relevant portion of the graph: the simulations we have reported here show that graphs with approximately three times more out-going than in-coming links in the most connected nodes exhibit the polylogarithmic scaling.
Establishing the exact connection between the in- and out-degree distributions and gap scaling is an interesting open problem for future research.

It would also be interesting to formulate a quantum circuit version of our PageRank algorithm. 
Perhaps the results obtained in \cite{HaHaLl} concerning the 
efficient solution of linear systems of equations could be used for this purpose.

{\it Acknowledgments}.---%
The authors are particularly grateful to Aram Harrow for insightful comments. Thanks also 
to Scott Aaronson, Richard Cleve, Hartmut Neven and Robert Kosut for useful discussions. 
This work was supported by NSF grants No. PHY-969969 and No. PHY-803304 to PZ and DAL. Additional support was provided to DAL by the NASA
Ames Research Center, the Lockheed Martin Corporation URI program, and the Google Research Award program.
 

\begin{thebibliography}{57}
\bibitem [{old()}]{old-PR-comment}%
  \BibitemOpen
  \href@noop {} {}\bibinfo {note} {Several recent papers have reported a
  physics approach to the properties of the Google matrix and the PageRank
  vector, e.g., \cite{Perra,PhysRevE.80.026107,PhysRevE.81.056109}. The focus
  of this research is mainly on localization phenomena occurring on the web
  graph, and unlike our work, it does not fully address computational
  complexity issues.}\BibitemShut {Stop}%
\bibitem [{com()}]{com-irred}%
  \BibitemOpen
  \href@noop {} {}\bibinfo {note} {An irreducible stochastic matrix implies
  that there exists a directed path from each node to any other
  node.}\BibitemShut {Stop}%
\bibitem [{gap()}]{gap-bound}%
  \BibitemOpen
  \href@noop {} {}\bibinfo {note} {This result only requires both $P_2^T$ and
  $E$ to be row-stochastic, and $E$ to have rank 1
  \cite{Nusbaum:03}.}\BibitemShut {Stop}%
\bibitem [{pag()}]{pagerank-comment1}%
  \BibitemOpen
  \href@noop {} {}\bibinfo {note} {When the initial Hamiltonian is chosen to be
  ``classical'', i.e., with a ground state equal to $|j\rangle$, the $j$th node
  of the web graph, instead of $h(G_c)$, our simulations show that it takes a
  time sublinear in $n$, though super-logarithmic, to generate the final ground
  state.}\BibitemShut {Stop}%
\bibitem [{sim()}]{sim-comment}%
  \BibitemOpen
  \href@noop {} {}\bibinfo {note} {{We used the Krylov subspace method
  \cite{Krylov} to propagate the Schr\"odinger equation subject to the
  Hamiltonian $h(s)$, and exact numerical diagonalization to extract the
  spectral gap.}}\BibitemShut {Stop}%
\bibitem [{sto()}]{stoq}%
  \BibitemOpen
  \href@noop {} {}\bibinfo {note} {We note that $H(s)$ is not stoquastic
  \cite{BrTe}, in that for all $s\in[0,1]$ it has both positive and
  negative off-diagonal matrix elements in the standard basis.}\BibitemShut
  {Stop}%
\bibitem [{DWa()}]{DWave-couplers-update}%
  \BibitemOpen
  \href@noop {} {}\bibinfo {note} {{Efficient $\Or(\log n)$ updating of
  coupling coefficients is indeed possible. For example, the Rainier chip
  developed by D-Wave Inc. supports such updating
  \cite{Johnson:10}.}}\BibitemShut {Stop}%
\bibitem [{gai()}]{gain}%
  \BibitemOpen
  \href@noop {} {}\bibinfo {note} {We have observed numerically that ${p_i}
  \propto 1/n^{\gamma_i}$, where $\gamma_i\in (0.6,1]$, and that $\Vert \vec{p}
  \Vert _2^2\propto 1/n$. Let $e_i$ denote the additive error corresponding to
  $\pi_i = p_i^2/\Vert \vec{p} \Vert_2^2 \sim n p_i^2$, and $e'_i$ the additive
  error corresponding to $p_i$. It follows from the Chernoff-Hoeffding
  inequality that the number of samples $M(x)$ from the distribution $x$, where
  $x=\pi=\{\pi_i\}$ (output of the quantum algorithm) or $x=\vec{p}=\{p_i\}$
  (PageRank) required for a given, fixed additive estimation error, is
  proportional to the inverse of the additive error: $M(\pi)\sim 1/e_i$,
  $M(\vec{p}) \sim 1/e'_i$. Assuming $e_i \sim \pi_i$ and $e'_i \sim p_i$ it
  follows that $M(\pi) \sim 1/\pi_i \sim 1/(n p_i^2) \sim n^{2\gamma_i-1}$,
  while $M(\vec{p}) \sim 1/p_i \sim n^{\gamma_i}$. The total cost required to
  prepare the sample in the quantum case is, as we have shown,
  $\Or[\mathrm{polylog}(n)]$, while it is at best $\Or[\log(n)]$ for classical
  MCMC for a stochastic matrix with constant gap, as is the case for the Google
  matrix. Thus, the fact that we have a speedup for estimation of the
  top-$\log(n)$ pages is due to the fact that we prepare the distribution $\pi$
  rather than $\vec{p}$. Of course, if the classical algorithm were modified to
  prepare $\pi$ rather than $\vec{p}$---presumably by an appropriate
  modification of $G$---the quantum algorithm could correpondingly be modified
  to prepare the distribution $\{\pi_i^2/\|\pi\|_2^2\}$, etc.}\BibitemShut
  {Stop}%
\end{thebibliography}

%

\appendix
\section{Supplemental material}
\subsection{Preferential attachment model and copying model}
The idea behind the preferential attachment model algorithm is that new vertices are more likely to attach to 
existing vertices with high degree. 
In our simulations we implemented the algorithm proposed in \cite{BoRiSp},
where some ambiguities of the original preferential attachment model \cite{BaAl} were resolved. 
This algorithm provides a scale-free network having a power-law degree distribution with 
a fixed exponent equal to $3$: $N(d)\propto d^{-3}$, where $N(d)$ is the number of nodes of degree $d$. 
A drawback of the  preferential attachment model is that 
global knowledge of the degree of all nodes is required. Moreover, the 
exponent of the power-law degree distribution is not controllable. 
The copying model introduced in \cite{KlKuRa} overcomes these drawbacks. 
It exploits only local structure to 
generate a power-law degree distribution. To do so
one starts from a small fixed initial graph of constant out-degree, and at each time step a 
pre-existing vertex is chosen uniformly at random. This node is called the copying vertex. For each 
neighbor of the copying vertex, a link is added 
from a new added vertex to that neighbor with probability $1-p$, while 
with probability $p$ a link is added from the new added 
vertex to a uniformly random chosen one. The parameter 
$p$ allows to obtain random graphs with power-law 
degree distributions with exponents given by ${(2-p)}/{(1-p)}$.

\subsection{Equivalence of non-local and local single-excitation Hamiltonians}

Here we show that the spectrum of the $N$-level Hamiltonian (acting on an $N$-dimensional Hilbert space)

\begin{equation}
h=\sum_{i=1}^{N} h_{ii} |i \rangle \langle i| +  \sum_{i < j} h_{ij} (|i \rangle \langle j| + |j \rangle \langle i|),
\label{eq:nonlocal}
\end{equation}
is the same as the spectrum of the following spin Hamiltonian (acting on the Hilbert space of $N$ qubits), when restricted to the single excitation manifold,

\begin{equation}
H = \sum_{i=1}^{N} h_{ii} \sigma^+_i \sigma^-_i + \sum_{i < j}^N h_{ij} (\sigma^+_i \sigma^-_j + \sigma^+_j \sigma^-_i)
\label{eq:local}
\end{equation}
where $ \sigma_k^{\pm} $ are Pauli ladder operators acting on the $k$th qubit. 
 
Since the Hilbert space of the $N$-qubit Hamiltonian is restricted to the single excitation manifold it is spanned 
 by $N$ basis vectors which can be put into one-to-one correspondence with the basis vector of the Hilbert space of the $N$-level Hamiltonian
 
 \begin{equation}
|i \rangle   \leftrightarrow |\uparrow \rangle_i \equiv| \downarrow_1 \cdots \downarrow_{i-1} \; \uparrow_i \; \downarrow_{i+1} \cdots \downarrow_N \rangle. 
\label{eq:map}
\end{equation}
Choosing the following representation for the Pauli matrices acting on qubit $j$,
\begin{align}
\sigma^x_j = |\uparrow \rangle \langle \downarrow|_j + |\downarrow \rangle \langle \uparrow|_j \notag \\
\sigma^y_j = i \left( |\downarrow \rangle \langle \uparrow|_j - |\uparrow \rangle \langle \downarrow|_j \right) \notag \\
\sigma^z_j = |\uparrow \rangle \langle \uparrow|_j - |\downarrow \rangle \langle \downarrow|_j \notag \\
\sigma^+_j =  \frac{\sigma^x_j+i\sigma^y_j}{2}=|\uparrow \rangle \langle \downarrow|_i \notag \\
\sigma^-_j = \frac{\sigma^x_j-i\sigma^y_j}{2}=|\downarrow \rangle \langle \uparrow|_i ,
\end{align}
one can derive Eq.~\eqref{eq:local} from Eq.~\eqref{eq:nonlocal} using Eq.~\eqref{eq:map}. 
The spectrum does not change in this construction since we are simply relabeling the bases 
of two isomorphic Hilbert spaces.  

Another way of seeing this is  to note that when the Hamiltonian in Eq.~\eqref{eq:local} is not restricted to the single excitation manifold 
and one has to diagonalize it, if the excitation numbers are conserved quantities, then one can first reduce the Hamiltonian into blocks labeled by the number of excitations and 
subsequently diagonalize each single block. The block labeled by a single excitation is equivalent to Eq.~\eqref{eq:nonlocal} via the mapping in Eq.~\eqref{eq:map}. 

\subsection{Role of the out-degrees}

The WWW graph is characterized by a power-law distribution for both for the in- and out-degrees of the nodes. 
Here we provide numerical evidence supporting the fundamental role played by the out-degrees in activating the polylogarithmic scaling of the 
average inverse gap, as a function of system size $n$ (the number of vertices in the graph). 

\begin{figure}
\includegraphics[scale=0.3]{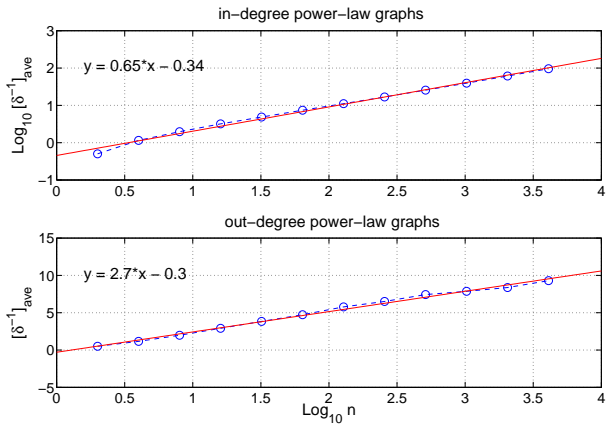} 
\caption{
(color online) 
Top panel: The average inverse minimum gap scaling for random graphs with only in-degree power-law distribution.  
Bottom panel: The average inverse minimum gap scaling for random graphs with only out-degree power-law distribution. 
1000 realizations.
}
\label{Fig:fig1SM}
\end{figure}

\begin{figure}
\includegraphics[scale=0.3]{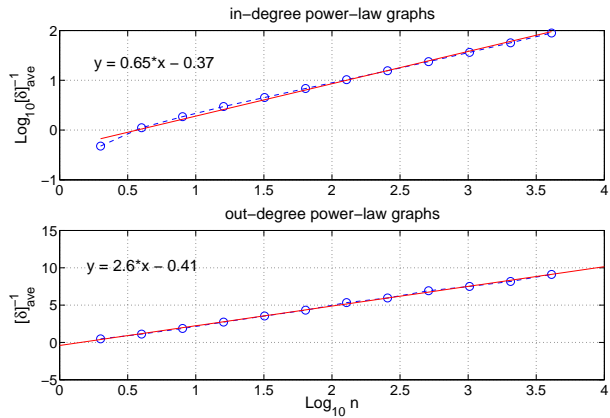} 
\caption{
(color online)
The inverse of the average minimum gap scaling for undirected preferential attachment random graphs. 
Top panel:  log-log plot.  
Bottom panel: semi-log plot. Linear fits are poor in both cases.
Averaged over 1000 realizations. 
}
\label{Fig:fig2SM}
\end{figure}

In order to distinguish the effect of the in-degrees from that of the out-degrees we consider preferential attachment
graphs constructed in such a way that only one  power-law is present. Starting with preferential attachment networks with only in-degree power-law 
distribution, Fig.~\ref{Fig:fig1SM} (top panel) shows the typical behavior of the inverse minimum gap. In this case the scaling is sub-linear, though not logarithmic: 
$[\delta^{-1}]_{\textrm{ave}} \sim n^{0.65}$. 
Also shown in Fig.~\ref{Fig:fig1SM} (bottom panel) is the scaling for the reverse graphs, obtained by reversing the direction of each edge. This corresponds to networks in which only the out-degrees are power-law distributed.
Remarkably, in this case we find the fit $ [\delta^{-1}]_{\textrm{ave}} \sim (\log_{10} n)^{2.7}$.
{In Fig.~\ref{Fig:fig2SM} we plot the same data considering the inverse of the average minimum gap, 
instead of the average of the inverse minimum gap. 
As expected qualitatively the scaling is the same, with small quantitative discrepancies.}

\begin{figure}
\includegraphics[scale=0.3]{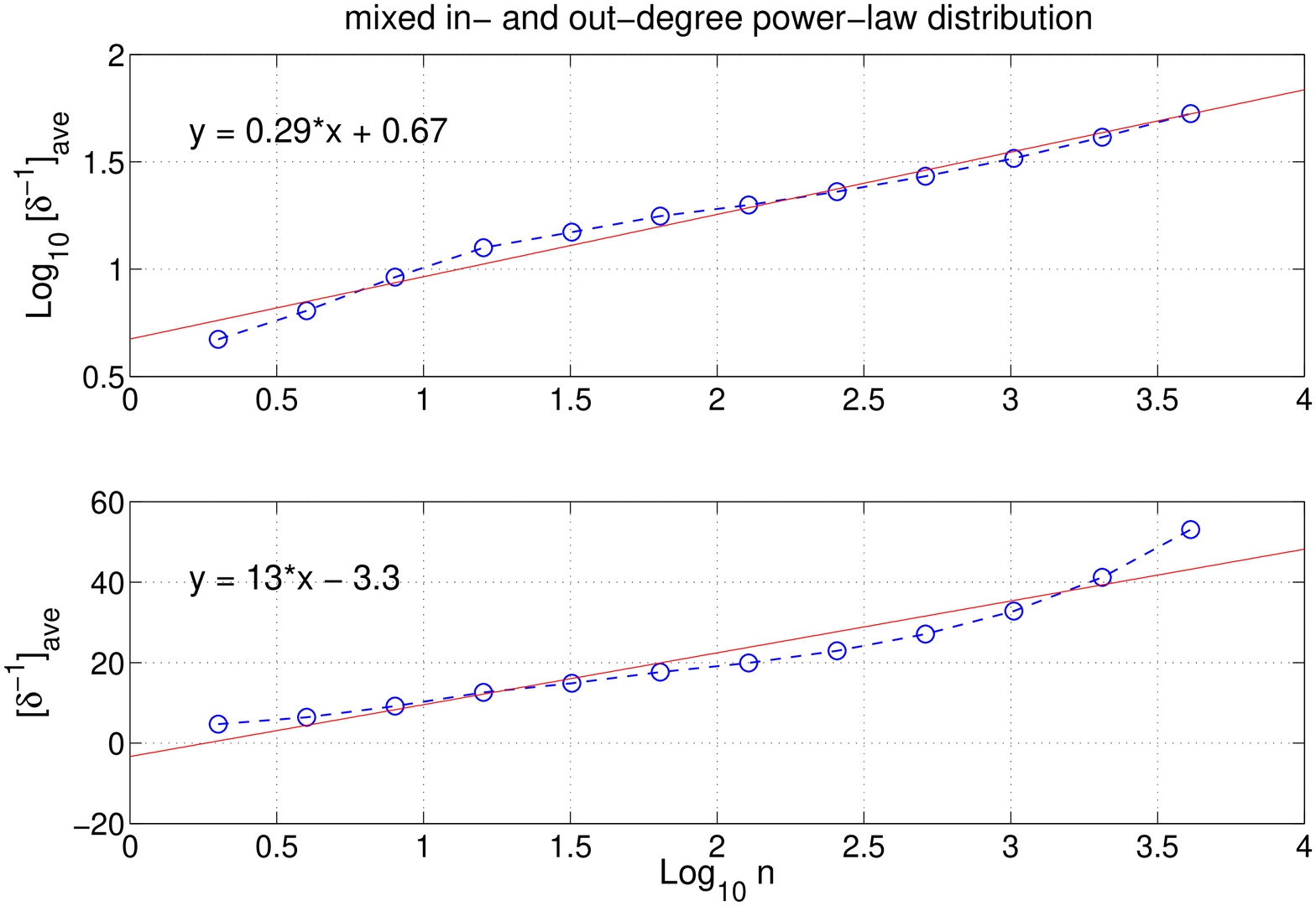} 
\caption{
(color online)
The average inverse minimum gap scaling for undirected preferential attachment random graphs. 
Top panel:  log-log plot.  
Bottom panel: semi-log plot. Linear fits are poor in both cases.
Averaged over 1000 realizations. 
}
\label{Fig:fig3SM}
\end{figure}

Fig.~\ref{Fig:fig3SM} shows what happens when we consider preferential attachment graphs with identical in- and out-degrees.
In this case the graph is equivalent to an undirected graph, and we find non-logarithmic, sub-linear scaling. 
We display both the double-logarithmic and the semi-logarithmic plots in order to make the distinction clear.

We note that the quantum adiabatic algorithm can still 
be useful even in the case of networks with only in-degree power-law distribution, for the preparation not of the pagerank state, but of the so-called \textit{inverse pagerank} \cite{GyGaPe} (used for spam detection). 
The latter is the pagerank of the  
reverse graph. The results of the simulations in Figs.~\ref{Fig:fig1SM}-\ref{Fig:fig3SM} suggest that, typically, when the algorithm is unable to prepare the pagerank in polylogarithmic time, it can still prepare the inverse pagerank in polylogarithmic time. 
\ignore{
The simulations shown in the main text were obtained by mixing (i.e., adding the adjacency matrices of) 
graphs $\mathcal{G}_A$, with only in-degree power-law distributions, with graphs 
$\mathcal{G}_B$ with only out-degree power-law distributions. 
For the family of graphs considered in the simulations reported in the main text, 
the maximum out-degree for $\mathcal{G}_B$ is approximately $3$ times greater than the maximum in-degree 
for $\mathcal{G}_A$. 
}
\end{document}